# EFFECT OF SYMMETRY ENERGY ON INTERMEDIATE MASS FRAGMENTS PRODUCTION

Rubina Bansal[*] and Suneel Kumar
*School of Physics and Material Science, Thapar University Patiala 147004, Punjab, INDIA*
*email: rubinabansal98@gmail.com

## Introduction

The science of nuclear physics deals with the properties of nuclear matter which makes up the massive centre of the atom. When two heavy nuclei collide at high energy, they create strongly interacting matter at energy densities far above that of normal nuclei. Depending upon the energy through which two nuclei collide different type of phenomena occurs. When energy of colliding nuclei is between 100-600 MeV/nucleon then multifragmentation take place. By studying the fragments (at final stage of reaction) we can seize the idea about initial condition and various other parameter which influence the reaction and vice-versa. The symmetry energy $E_{\text{sym}}$ $(\rho)$ of nuclear matter characterizes how the energy rises as one move away from equal numbers of neutrons and protons. Both the magnitude and density dependence of $E_{\text{sym}}$ $(\rho)$ are critical for understanding the structure of rare isotopes and the reaction mechanism of heavy ion collisions but also many interesting issues in astrophysics. The best estimate of the density dependence of the symmetry energy can be parameterized as[3]

$$C_{\text{sym}}(\rho) = C_{\text{sym}}^0 \left(\frac{\rho}{\rho_0}\right)^\gamma \text{ (MeV)}$$

where $C^o_{\text{sym}}$, is the value of the symmetry energy at normal density and $\gamma$ is the parameter that characterizes the stiffness of the symmetry energy. Here we correlate the symmetry energy with cross section and see the effect on production of intermediate mass fragments (IMF) of Au+Au at different incident energies.

## IQMD model

We use Isospin dependent Quantum Molecular Dynamics (IQMD) model [1] to see the effect of cross-section and symmetry energy on fragments production. IQMD is extended form of quantum molecular dynamics (QMD) [2]. In this model the baryons are represented by Gaussian shaped density distributions

$$f_i(\vec{r},\vec{p},t) = \frac{1}{\pi^2 \hbar^2} e^{-[\vec{r}-\vec{r}_i(t)]^2 \frac{1}{2L}} e^{-[\vec{p}-\vec{p}_i(t)]^2 \frac{2L}{\hbar^2}}$$

The Hamilton equations of motion for the propagation of hadrons are

$$\dot{r}_i = \frac{\partial H}{\partial p_i}; \qquad \dot{p}_i = -\frac{\partial H}{\partial r_i}$$

Where H stands for the Hamiltonian which is given by:

$$H = \sum_{i,j}^{A} \frac{p_i^2}{2m} + \sum_{i}^{A}(V_{ij}^{Sk} + V_{ij}^{Yuk} + V_{ij}^{Cou} + V_{ij}^{mdi} + V_{ij}^{sym})$$

Where $V_{ij}^{Sk}$, $V_{ij}^{Yuk}$, $V_{ij}^{Cou}$, $V_{ij}^{mdi}$, and $V_{ij}^{sym}$ are the Skyrme, Yukawa, Coulomb, momentum dependent interaction (MDI), and symmetry potentials respectively.

## Results and discussion

In the following discussion, we have simulated the 1000 events for $_{79}Au^{197} + _{79}Au^{197}$ at incident energies 50, 100, 200, 400, 600, 1000 MeV/nucleon using different collision geometries with different form of density dependent symmetry energy. For the analysis, soft as well as momentum independent(SMD) equation of state with compressibility K = 200 MeV is used. If matter is highly compressed, the nucleon-nucleon correlations are broken due to violent nucleon-nucleon collision. We see the effect of energy, symmetry energy and cross section on the production of IMF at scaled impact parameter $\hat{b}$ = 0.3.

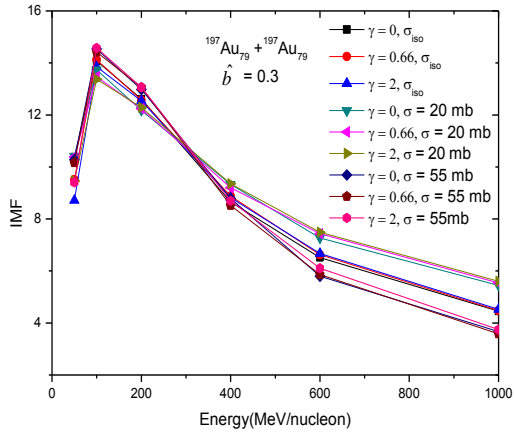

*Figure.1: Effect of incident energy, symmetry energy and cross section on IMF production .*

From figure 1, we observed the multiplicity of IMF's is maximal at energy 100 MeV/nucleon then it decrease with increase the energy. Due to the low excitation energy, central collisions generate better repulsion and break the colliding nuclei into IMF's. One can see the global universality of rise and fall in the multiplicity of the IMF's production. In all the cases, there are very little effects of different cross sections at 100 MeV/nucleon and effect of cross section increase at higher energies. In the other words, we can say that the production of IMF's effected by different cross section at higher energies.

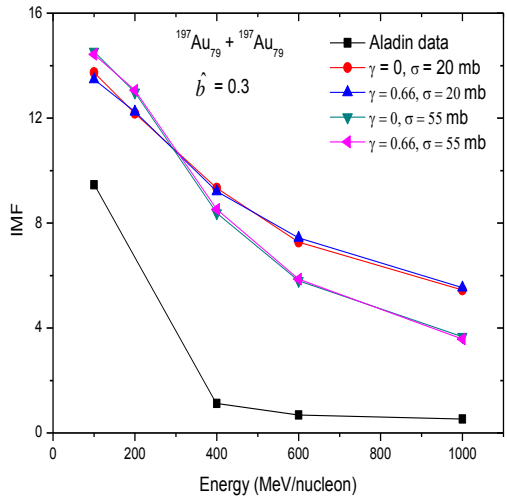

*Figure 2: Comparison of average multiplicity of intermediate mass fragments (IMF's) with ALADIN data at incident energies of 100, 200, 400, 600, 1000 MeV/nucleon as a function of energy.*

In Fig.2, we are comparing our results i.e. production of IMF's with experimental data of ALADIN of $^{197}Au_{79}+^{197}Au_{79}$ at different energies at fixed cross sections. We observed that the production of IMF according to our calculations and experimental data shows similar kind of rise and falls. For central geometry the collisions are violent so there are few numbers of IMF's observed. At peripheral collisions maximum part of target and projectile goes unintracted, resulting into heavy mass fragments. Thus very few intermediate mass fragements.

## References

[1] C. Hartnack et al., Eur. Phys. J. **A 1**, 151(1998); S.Kumar, S. Kumar and R. K. Puri, Phys.Rev. **C 81**, 014611(2010); V. Kaur, S.Kumar and R. K. Puri, Phys. Lett. **B 697**, 512(2011);V. Kaur, S. Kumar and R. K. Puri, Nucl. Phys. **A 861** , 37(2011).

[2] J. Aichelin, Phys. Rep. **202**, 233(1991); R. K. Puri, et al., Nucl. Phys. **A 575**, 733(1994); ibid. J. Comp. Phys. **162**, 245(2000); E. Lehmann, R. K. Puri, A. Faessler, G. Batko, and S.W. Huang, Phys. Rev. **C 51**, 2113(1995); ibid. Prog. Part. Nucl. Phys. **30**, 219(1993); Y. K. Vermani et al., J. Phys. G. Nucl. Part. Phys. **36**, 0105103(2009); ibid **37**, 015105(2010); ibid Phys. Rev**. C 79**, 064613(2009); ibid Nucl. Phys. **A 847**, 243(2010). S. Kumar, S. Kumar and R. K. Puri, Phys. Rev. **C 78**, 064602; ibid **57**, 2744; S. Goyal and R. K. Puri, Phys. Rev. **C 83**, 47601.

[3] L. W. Chen, C.M. Ko, andB. A. Li,Phys. Rev. Lett. **94**, 032701 (2005).